\documentclass[prl,twocolumn,letterpaper,showpacs]{revtex4}

\usepackage{amssymb}
\usepackage{epsfig}

\begin{document} 

\title{Perfect quantum state transfer with spinor bosons on weighted graphs}

\author{David~L.~Feder} 
\affiliation{Department of Physics and Astronomy and Institute for Quantum
Information Science, University of Calgary, Calgary, Alberta, Canada T2N 1N4}

\date{\today}

\begin{abstract}
A duality between the properties of many spinor bosons on a regular lattice
and those of a single particle on a weighted graph reveals that a quantum
particle can traverse an infinite hierarchy of networks with perfect
probability in polynomial time, even as the number of nodes increases
exponentially. The one-dimensional `quantum wire' and the hypercube are
special cases in this construction, where the number of spin degrees of
freedom is equal to one and the number of particles, respectively. An
implementation of near-perfect quantum state transfer across a weighted
parallelepiped with ultracold atoms in optical lattices is discussed.
\end{abstract}


\maketitle

The quantum walk (QW) is the quantum mechanical extension of the classical
random walk, where the quantum particle (the walker) can be thought of as
following many trajectories simultaneously~\cite{Reviews}. For quantum
walks on regular (unweighted) linear graphs, the mean displacement of the
particle is quadratically faster than that of a classical random
walk~\cite{Aharonov01}. The time needed to propagate from the input to the
output vertex for certain two-dimensional regular graphs with randomization
has been shown to be exponentially faster than with any classical
algorithm~\cite{Childs02}, but like the linear graph the probability of
hitting the output vertex decreases polynomially with the graph width. Perfect
output probability can be achieved on the $n$-dimensional
hypercube~\cite{Shenvi03,Kempe05} based on regular and weighted
one-dimensional graphs~\cite{Christandl04} with an exponential speed-up over a
classical random walk, though the improvement is only polynomial compared with
a suitable classical algorithm~\cite{Childs02}.

%The fast spreading of quantum walks in principle provides a powerful approach
%to the construction of quantum algorithms that outperform their classical
%counterparts. To date, these include quantum search
%algorithms~\cite{Shenvi03,Childs04,Tregenna03,Ambainis05}, oracular
%(black-box) problems which show exponential speed-up over classical
%methods~\cite{Childs03}, element distinctness~\cite{Ambainis04}, the
%triangle problem~\cite{Magniez05}, and subset finding~\cite{Childs05}.

Quantum walks on weighted graphs have been proposed as an efficient way
to transfer quantum states (and therefore quantum information) with perfect
fidelity without requiring external control~\cite{Christandl04}. To maximize
throughput, it would be most convenient to transport many quantum states
simultaneously. Unfortunately, the quantum particles on which the information
is encoded are usually \emph{indistinguishable}; if the receiver needs to
know both the information and the particular carrier (where the latter might
represent a given wire in a quantum circuit, for example), then the particles
need to be spatially or temporally separated. For many applications it is
sufficient to know only that a certain amount of information has been
transmitted, however. For example, Alice might want to send Bob $M$ copies of
a given quantum state (entangled or otherwise), or one qudit each of $M$ Bell
pairs so that Alice and Bob can share $M$ ebits, or $N-1$ qudits of a GHZ$_N$
or complete graph state, etc.

As discussed in detail below, many spinor bosons undergoing continuous-time
quantum walks (CTQWs) can be mapped to a single particle walking on an
infinite hierarchy of weighted graphs, including (but not
limited to) weighted hyperparallelepipeds, hypertetrahedra and hyperoctahedra.
These graphs generally have vertices of variable degree, and share the
property with the hypercubes that they can be traversed with unit probability
in polynomial time. Unlike the hypercubes, however, there is no known
classical algorithm that can accomplish the same task. The unique properties
of walks on these graphs may aid in the development of new schemes for quantum
communication and computation.

Consider $N$ bosons with $S$ spin (or pseudospin) degrees of freedom, located
on a connected graph with $V$ vertices of minimal degree one. The
$(N_{\sigma}+V-1)!/N_{\sigma}!/(V-1)!$ accessible quantum states for each spin
projection $\sigma\in\{-{(S-1)\over 2},-{(S-3)\over 2},\ldots,{(S-3)\over 2},
{(S-1)\over 2}\}$ are defined by the Bose occupation of the
various sites; that is, the natural basis states are defined in Fock
(occupation number) space. Explicitly, for a two-site lattice the
$\prod_{\sigma}(N_{\sigma}+1)$
states are
\begin{equation}
|n_1,n_2\rangle=\prod_{\sigma\sigma'}{1\over\sqrt{n_{1\sigma}!n_{2\sigma'}!}}
\left(a_{1\sigma}^{\dag}\right)^{n_{1\sigma}}\left(a_{2\sigma'}^{\dag}
\right)^{n_{2\sigma'}}|{\bf 0}\rangle,
\end{equation}
where the quantum numbers $n_{i\sigma}\in\{0,1,2,\ldots,N\}$ denote the
number of bosons with spin projection $\sigma$ in the site $i$, $|{\bf
0}\rangle$ denotes the particle vacuum, and $\sum_{i\sigma}n_{i\sigma}=N$. In
this second-quantized notation, the
operators $a^{\dag}_{i\sigma}$ and $a^{\vphantom{\dag}}_{i\sigma}$ create and
annihilate, respectively, a boson with spin projection $\sigma$ in site $i$.
The normalization factors reflect the $n_{i\sigma}!$ ways that $n_{i\sigma}$
identical bosons can be arranged among themselves. Thus each basis state
encodes the symmetric permutation of all quantum states with the same
Hamming weight. The basis states for $N$ bosons on $V$ sites are obvious
generalizations of these, requiring at most $VS$ quantum numbers
$n_{1\sigma},n_{2\sigma},\ldots,n_{V\sigma}$.

The dynamics of this system are described by the hopping Hamiltonian:
\begin{equation}
H=\sum_{\langle ij\rangle\sigma}\tau_{ij}\hat{a}_{j\sigma}^{\dag}
\hat{a}_{i\sigma}^{\vphantom{\dag}},
\label{Ham}
\end{equation}
where $\tau_{ij}$ is amplitude to hop between sites $i$ and $j$, and the angle
brackets denote a sum over only nearest neighbors of a given vertex
connected by an edge. The
hopping amplitudes therefore define the adjacency matrix of the graph,
which is assumed to be undirected ($\tau_{ij}=\tau_{ji}$) but otherwise
unconstrained. The only non-zero matrix elements of the Hamiltonian~(\ref{Ham})
connect an $N$-particle Fock basis state to another where one boson with spin
projection $\sigma$ is annihilated at site $i$ and another with the same spin
is created in an adjacent site $j$. Consider, for example, the matrix element
connecting the states $|n_1,n_2,n_3\rangle$ and $|n_1,n_2-1,n_3+1\rangle$,
assuming all spin states are equal. Because each quantum number labels a
different (site) subspace, the matrix element decomposes into a tensor product
of matrix elements for the three sites. If the hopping amplitude between these
states is $\tau$ one obtains:
\begin{eqnarray}
&&\langle n_1,n_2-1,n_3+1|H|n_1,n_2,n_3\rangle\nonumber \\
&&\quad=\tau{1\over n_1!}\langle 0|\left(a_1^{\vphantom{\dag}}\right)^{n_1}
\left(a_1^{\dag}\right)^{n_1}|0\rangle\nonumber \\
&&\qquad\otimes{1\over\sqrt{(n_2-1)!n_2!}}\langle 0|
\left(a_2^{\vphantom{\dag}}\right)^{n_2-1}a_2
\left(a_2^{\dag}\right)^{n_2}|0\rangle\nonumber \\
&&\qquad\otimes{1\over\sqrt{(n_3+1)!n_3!}}\langle 0|
\left(a_3^{\vphantom{\dag}}\right)^{n_3+1}a_3^{\dag}
\left(a_3^{\dag}\right)^{n_3}|0\rangle\nonumber \\
&&\quad=\tau\sqrt{n_2!(n_3+1)!\over(n_2-1)!n_3!}=\tau\sqrt{n_2(n_3+1)}.
\label{adjacency}
\end{eqnarray}
All other elements of the Hamiltonian can be obtained by suitably replacing
any of the $n_i$. 

The $N$-particle quantum walk on the original (primary) graph is therefore
dual to a one-particle walk on a secondary graph whose vertices are labeled by
the Fock states and whose adjacency matrix is defined by the Hamiltonian matrix
elements above. Since the Hamiltonian is constant in time, the CTQW walk is
effected by propagating the Fock-space wavefunction according to
$|\psi(t)\rangle=U(t)|\psi(0)\rangle$ where the time-dependent unitary is
$U(t)=e^{-iHt}$. Starting with unit probability
on the input vertex, the CTQW is executed until the probability at the output
vertex is maximized; this defines the hitting time $t_h$. The choice of
input and output vertices depends on the structure of the graph, and will be
clarified below.

Suppose that the primary graph $P$ is the two-vertex chain, with the input and
output vertices corresponding to left and right sites, respectively; this can
be traversed with 100\% probability with a CTQW in constant time
$t_h=(\pi/2)\tau^{-1}$~\cite{Shenvi03}. This value of $t_h$ is independent of
$N$ because each boson executes its own independent quantum walk on the
primary graph. The original eigenvalues $\lambda=\{\pm\tau\}$ are transformed
into the $N$-multiples $\lambda=\tau\{-N,-N+2,\ldots,N-2,N\}$ which are always
commensurate~\cite{Christandl04}. The resulting structure of the evolution
operator $U(t)$ on the secondary graph ensures that perfect state transfer
occurs between each pair of symmetry-related vertices at $t=t_h$, i.e.\
between any two vertices whose quantum numbers are related by inversion
$|n_{1\sigma},n_{2\sigma},\ldots,n_{V\sigma}\rangle\leftrightarrow|n_{V\sigma},
\ldots n_{2\sigma},n_{1\sigma}\rangle$.

Choosing $N=S$
so that there are $N$ distinguishable bosons, then $n_{i\sigma}\in\{0,1\}$ for
each value of $\sigma$ and the matrix elements~(\ref{adjacency}) are all equal
to $\tau$. As expected, the system maps to the $N$-dimensional $\tau$-length
hypercube $P^{\otimes N}$, which has the same value of $t_h$. When $S=1$ so
that all the bosons are spinless and therefore indistinguishable, the secondary
graph $G_N$ is the weighed line whose $N+1$ vertices are labeled by the Fock
states $|N-n,n\rangle$ where $n\in\{0,1,\ldots,N\}$ and the edge weights are
given by $H_{n,n+1}=\tau\sqrt{(n+1)(N-n)}$. These are precisely the weights
required to transport a quantum state across a quantum wire, also in time 
$t_h=(\pi/2)\tau^{-1}$~\cite{Christandl04} which has been shown to be
optimal~\cite{Yung06}. In fact, as has been noted~\cite{Shenvi03}, these
weights result from projecting the vertices of an $N$-dimensional unit-length
hypercube onto a line. The boson
construction indicates that the projection is driven by indistinguishability:
all states of the hypercube with the same Hamming weight are mapped to a
common point on the line, but the bit positions in the original strings are
not specified.

Suppose that the $N$ bosons each carry the qudit state 
$|\psi\rangle=c_1|-{(S-1)\over 2}\rangle+\ldots+c_S|{(S-1)\over 2}\rangle$,
where $\sum_{i=1}^S|c_i|^2=1$. Because the Hamiltonian~(\ref{Ham}) preserves
the total spin, the $N$-particle walk decomposes into a superposition of
walks on $S$ graphs of the type $G_N$ (quantum wires), $S(S-1)$ graphs of type
$G_{N-1}\otimes P$ (weighted rectangles), $S(S-1)(S-2)/2$ graphs of type
$G_{N-2}\otimes P^{\otimes 2}$, $S(S-1)$ graphs of type $G_{N-2}\otimes G_2$,
etc. These secondary graphs are hyperparallelepipeds of maximum dimension $S$,
where along each direction the edge weights correspond to those of the quantum
wire. Perfect quantum state transfer is effected between any two vertices
connected by a line through the geometric center of the hyperparallelepiped.

Consider now $N'$ bosons randomly distributed on one of the weighted secondary
graphs. At $t=t_h$, the quantum state labeled by the Bose
occupations for each site is transformed into another state with the same
populations in the symmetry-related sites. This final quantum state
corresponds to a site that is itself the symmetry counterpart to the initial
site on a tertiary graph. In this manner one can generate an infinite hierarchy
of weighted graphs that all share the properties of the primary graph. The
largest edge weight is proportional to $N_m\cdots N_2N_1$ where $N_m$ is the
number of bosons at recursion level $m$; rescaling this weight back to $\tau$
yields a physical hitting time $t_h\propto\prod_{i=1}^mN_i$
that increases polynomially even as the number of vertices grows exponentially.

\begin{figure}[t]
\epsfig{file=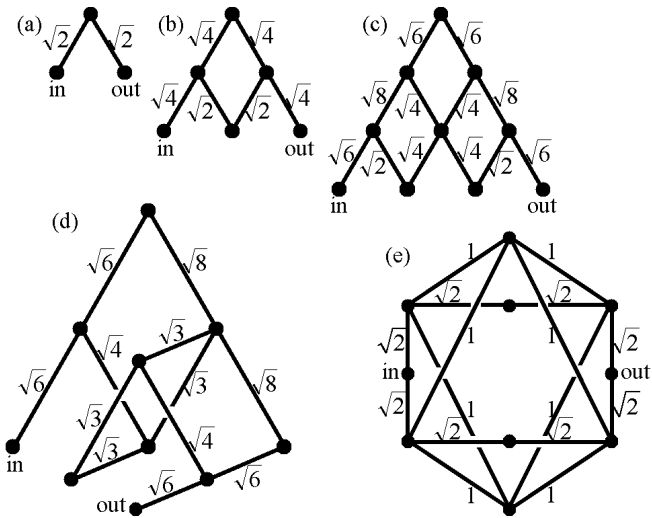,width=\columnwidth,angle=0}
\caption{Examples of small tertiary graphs built from the quantum wire and
hypercube. Cases (a) through (c) correspond to $N=1$ through $3$ on $G_2$;
cases (d) and (e) depict $N=2$ on $G_3$ (tetrahedron of length 2) and on
$P\otimes P$ (octahedron of base length 2), respectively. All weights are
in units of $\tau$. A CTQW running for $t=(\pi/2)\tau^{-1}$ will yield perfect
quantum state transfer between vertices `in' and `out'.}
\label{hyper}
\end{figure}

The smallest tertiary graphs built from quantum wires and hypercubes are shown
in Fig.~\ref{hyper} for $S'=1$. With $N'$ indistinguishable bosons on $G_N$,
the resulting graphs are $N$-dimensional weighted hypertetrahedra with $N'+1$
vertices along each edge, but where not all vertices share an edge with their
nearest neighbors. For the two-dimensional graphs defined by $N'$ bosons
on $G_2$, shown in Figs.~\ref{hyper}(a-c), the edges between $V$ vertices
along a given constant direction have weights defined by $V-1$ bosons on $P$,
multiplied by $\sqrt{2}$; perfect quantum transfer occurs between any two 
vertices related by a reflection about the graph center, so that states on
central vertices remain fixed at intervals of $t_h$. $N'$ bosons on the square
$\square$ (two-dimensional unweighted hypercube) map to a single boson on
(generally truncated) octahedra with $N'+1$ vertices along each edge of the
square base, as shown in Fig.~\ref{hyper}(e); quantum states are transferred
to between vertices through the body center and parallel to the base, so that
states on the two apices remain fixed at $t_h$ intervals.

The edge weights on high-order graphs in this construction generally increase
both with $N'$ and the number of vertices, as is evident in
Figs.~\ref{hyper}(a)-(c). Consider for example the graphs defined by a
hierarchy of spinless bosons
$\{N_m\;\mbox{on}\;[\ldots N_1\;\mbox{on}\;(N_0\;\mbox{on}\;2^{\otimes d})]\}$,
where $d$ is the secondary hypercube dimension and $m$ is the number of
iterations in the hierarchy used to construct the graph. Because of the 
symmetry of the underlying graphs at each level, edges with the largest 
weights are contiguous. This implies an efficient classical algorithm to
traverse these graphs with high probability and relatively few steps: one
simply chooses the `heaviest' edges at each step. 

\begin{figure}[t]
\epsfig{file=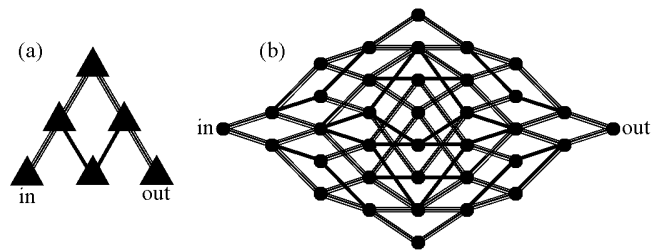,width=\columnwidth,angle=0}
\caption{Two representations of the same four-dimensional fourth-order graph
$(2\;\mbox{on}\;G_2)^{\otimes 2}=[2\;\mbox{on}\;(2\;\mbox{on}\;2)]^{\otimes 2}$.
Each triangular node in (a) corresponds to the six-vertex weighted graph shown
in Fig.~\ref{hyper}(b); each node within a triangle connects to its counterpart
in an adjacent triangle by an edge with the same weight. Solid and striped
lines correspond to edges with weight $\sqrt{2}$ and $\sqrt{4}$, respectively.}
\label{graphs2}
\end{figure}

Graphs avoiding this issue include constructions such as
$\{N_m\;\mbox{on}\;\ldots [N_1\;\mbox{on}\;(N_0\;\mbox{on}\;2)]^{\otimes S_1}
\ldots \}^{\otimes S_m}$, with $N_1$ distinguishable bosons on hypertetrahedra
forming the underlying graph for $N_2$ distinguishable bosons, etc. An example
is shown in Fig.~\ref{graphs2}. When these graphs get very large, the weights
tend to become homogeneously distributed; because the vertex degrees also vary
across the network, there are no identifiable criteria that can be used to
classically determine the optimal edge at each step. Together with an
exponential increase in the number of vertices, this implies an exponential
speed-up of the quantum walk over any classical traversal algorithm.

The two-vertex graph need not be the only primary graph that yields perfect
quantum state transfer, however, if the weights are allowed to be complex. The
resulting hierarchy of higher-order graphs are the quantum analog of signed
classical networks. In this case the hopping Hamiltonian becomes
$H=\sum_{\langle ij\rangle\sigma}(\tau_{ij}\hat{a}_{j\sigma}^{\dag}
\hat{a}_{i\sigma}^{\vphantom{\dag}}+\mbox{H.c.})$, where $\tau$
is now complex and `H.c.' stands for `Hermitian conjugate.' One of these is the
triangle $\triangle$ with edge weights $i\tau/\sqrt{3}$ and eigenvalues
$\lambda=\{\pm\tau,0\}$. Labeling the vertices clockwise around $\triangle$,
the state is transfered to each successive neighbour in a counterclockwise
fashion after each time interval $t=(2\pi/3)\tau^{-1}$. The secondary graphs
generated with $N$ bosons on $\triangle$ correspond to weighted triangles and
tetrahedra. These are much like Figs.~\ref{hyper}(a-c), in that the weights
are defined by the $G_V$ graph with $V+1$ vertices along a given direction
(multiplied by $i/\sqrt{3}$), except that all nearest neighbors now share an
edge. Quantum states are successively transferred between vertices
related by a counterclockwise $2\pi/3$ rotation about the center, which is a
symmetry operation for the graph. Another example is the square $\square$ or
bowtie $\bowtie$ with successive
weights $1,e^{i\alpha},e^{i\beta},e^{i\gamma}$ in units of $\tau$, where the
first weight can be chosen to be real without loss of generality; $P\otimes P$
is a special case with $\alpha=\beta=\gamma=0$. Whenever $\gamma=\alpha+\beta$,
the eigenvalues are $\lambda=\{\pm\tau\}$ and perfect state transfer occurs
between vertices at opposite corners.

That the eigenvalues of the hopping Hamiltonian on the primary graph are
commensurate and symmetric about zero is
a necessary but insufficient condition for perfect state transfer, however.
For example, the complete four-vertex graph can have weights $a$ on outside
edges and weights $b$ on the inside (diagonal) edges, while still preserving
$\pi/4$-rotation symmetry. For example, choosing $a=i/\sqrt{2}$ and $b=i$ yields
$\lambda=\{\pm 2\tau,0\}$ but after $t_h=\pi/2\tau$ the state is in an equal
superposition of two vertices.
%; likewise when choosing $a=1/2$ and
%$b=i\sqrt{2}$, which yields $\lambda=\{\pm\tau,\pm 2\tau\}$.
While it is conceivable that a judicious choice of weights could guarantee
perfect state transfer between two nodes of an arbitrary graph, the
prescription for accomplishing this has not been determined.

Ultracold atomic spinor bosons confined in optical
lattices~\cite{Stoferle04,Paredes04,Fertig05} provide a physical system
that is particularly amenable to implementing perfect quantum state transfer
on some of the weighted networks discussed above, especially those with
regular geometries. The atomic interactions can be made small through the use
of Feshbach resonances~\cite{Cornish00}, and the tunneling amplitudes and
phases from site to site can be controlled to some extent~\cite{Mueller04}.
Particularly simple graphs to physically
construct are the secondary weighted parallelepipeds $G_V^{\otimes d}$ where
$1\leq d\leq 3$ and the weighted tetrahedra based on $\triangle$ with
complex weights. The
tunneling matrix elements along any given direction now depend on position,
$\tau_{i-1,i}=\tau\sqrt{i(V-i+1)}$ with $1\leq i\leq V$. With the substitution
$i=[(V+1)/2]-j$ where $j$ is the vertex index relative to the graph center,
one finds $\tau_{j-1,j}^2\propto[(V+1)/2]^2-j^2$.
For standard optical lattice potentials of the form
$\hat{V}(\tilde{x})=sE_R\cos^2(k\tilde{x})$, where $\tilde{x}=\{x,y,z\}$,
$k=2\pi/\lambda$ is the wavevector ($\lambda$ is the laser wavelength), and
$s$ is the lattice depth in units of the atomic recoil energy
$E_R=\hbar^2k^2/2m$ ($m$ is the atomic mass), the tight-binding approximation
applicable for $s\gtrsim 5$ gives $\tau^2=(16/\pi)s^{3/2}e^{-4\sqrt{s}}$ in
recoil energies~\cite{Pupillo06}. The required quadratic
variation of $\tau_{j-1,j}^2$ can be approximately obtained by choosing a weak
position-dependent lattice depth $s=s_0+aj^2$ in each direction, which in
principle can be accomplished by appropriately focusing the lattice laser beam
and adding end caps to provide hard-wall boundary conditions.

The Schr\" odinger equation for bosons in a one-dimensional (1D) optical
lattice with $V=101$ and $s_0=5$ was propagated in time numerically using a
finite-element discrete variable approach~\cite{Schneider06}. Choosing
$a=10^{-3}$ and $S=1$, an initial quantum state centered at
$j_0=\{10,20,30,40\}$
and spread over several sites was found to transfer to the sites located
symmetrically opposite the lattice center with probabilities exceeding 99\%. 
The hitting times were within 10\% of the theoretical value
$t_h=(V+1)(\pi^{3/2}/16)s^{-3/4}e^{2\sqrt{s}}(\hbar/E_R) \approx 40$~ms,
assuming $^{87}$Rb atoms in a lattice with $\lambda=800$~nm. Surprisingly, the
initial wavepacket remains well-localized during the propagation. Weak damping
of the oscillations between initial and final states is due to increased
dispersion driven by the imperfect phase profile (which is more pronounced
near the lattice edge), and was found to have a characteristic time between
$42t_h$ ($j_0=40$) and $116t_h$ ($j_0=10$).

Precise control of the lattice potential or atomic interactions is not required
to effect the state transfer. For example, choosing $a=7\cdot 10^{-3}$ when
$V=101$ gives a variation of the hopping amplitudes that is more appropriate
for a lattice with $V=41$. State transfer with probability exceeding 95\% was
found for any $j_0\leq 20$ within approximately 10\% of $t_h=380\tau^{-1}$. For
$j_0>20$, the wavepacket remained at the lattice periphery. Virtually
identical behavior was obtained using a Gaussian, rather than quadratic,
profile for the lattice height. Weak repulsive interactions between
bosons were incorporated in a mean-field sense by solving the appropriate
quasi-1D nonlinear Schr\" odinger equation. The strength of the additional
term $g|\psi|^2$ is parametrized by the effective 1D interaction strength $g$,
which in the current
dimensionless units is $g=a_s\lambda/\pi^2\ell_{\perp}$~\cite{Olshanii98}
where $a_s$ is the $s$-wave scattering length and $\ell_{\perp}$ is the
transverse oscillator length. For parameters relevant to current
experiments~\cite{Fertig05}, $g\approx 0.2$. For values of $g\leq 1$, $t_h$
is found numerically to increase only slightly (by 3\%), and the probability
of hitting the output vertex decreases to 95\%. Because the $t_h$ are
identical for all sites, the behavior for this range of interaction strengths
is only weakly dependent on the initial density distribution. For larger
values of $g$ the output probability drops precipitously, but the mean-field
approximation becomes increasingly suspect. The numerical results suggest that
near-perfect quantum state transfer can be achieved using ultracold bosons in
weighted optical lattices under realistic experimental conditions.

\begin{acknowledgments}
It is a pleasure to thank Peter H\o yer and John Watrous for stimulating
conversations. This work was supported in part by the Natural Sciences and
Engineering Research Council of Canada and the Canada Foundation for
Innovation.
\end{acknowledgments}


\begin{thebibliography}{99}

\small 

\bibitem{Reviews}
J.~Kempe, {\it Contemporary Physics}~{\bf 44} 307 (2003); A.~Ambainis,
Int. J. Quantum Inf.~{\bf 1}, 507 (2003).

\bibitem{Aharonov01}
D.~Aharonov, A.~Ambainis, J.~Kempe, and U.~Vazirani,
Proceedings of ACM Symposium on the Theory of Computation, 50 (2001).

\bibitem{Childs02}
A.~M.~Childs, E.~Farhi, and S.~Gutmann, Quantum Inf. Proc.~{\bf 1}, 35 (2002).

\bibitem{Shenvi03}
N.~Shenvi, J.~Kempe, and K.~B.~Whaley, Phys. Rev. A~{\bf 67}, 052307 (2003).

\bibitem{Kempe05}
J.~Kempe, Probability Theory and Related Fields~{\bf 133}, 215 (2005).

\bibitem{Christandl04}
M.~Christandl, N.~Datta, A.~Ekert, and A.~J.~Landahl, Phys. Rev. Lett.~{\bf 92},
187902 (2004); M.~Christandl {\it et al.}, Phys. Rev. A~{\bf 71}, 032312 (2005).

%\bibitem{Childs04}
%A.~M.~Childs, and J.~Goldstone, Phys. Rev. A~{\bf 70}, 022314 (2004).

%\bibitem{Tregenna03}
%B.~Tregenna, W.~Flanagan, R.~Maile, and V.~Kendon, New J. Phys.~{\bf 5}, 83
%(2003).

%\bibitem{Ambainis05}
%A.~Ambainis, J.~Kempe, and A.~Rivosh, Proceedings of the 16th ACM-SIAM
%Symposium on Discrete Algorithms, 1099 (2005). 

%\bibitem{Childs03}
%A.~M.~Childs {\it et al.}, Proceedings of ACM Symposium on Theory of
%Computation, 59 (2003).

%\bibitem{Ambainis04}
%A.~Ambainis, Proceedings of the 45th Annual IEEE Symposium on Foundations of
%Computer Science, 22 (2004).

%\bibitem{Magniez05}
%F.~Magniez, M.~Santha, and M.~Szegedy, Proceedings of the 16th ACM-SIAM
%Symposium on Discrete Algorithms, 1109 (2005). 

%\bibitem{Childs05}
%A.~M.~Childs and J.~M.~Eisenberg, Quantum Inf. and Comput.~{\bf 5}, 593 (2005).

%\bibitem{Brun05}
%H.~Krovi and T.~Brun, quant-ph/0510136.

\bibitem{Yung06}
M.-H.~Yung, quant-ph/0603179.

%\bibitem{Moore02}
%C.~Moore and A.~Russell. Quantum walks on the hypercube. Proceedings of RANDOM
%02, 164 (2002).

%\bibitem{noh04}
%J.~D.~Noh, and H.~Rieger,Phys. Rev. Lett.~{\bf 92}, 118701 (2004).

\bibitem{Stoferle04}
T.~St\" oferle {\it et al.}, Phys. Rev. Lett.~{\bf 92}, 130403 (2004).

\bibitem{Paredes04}
B.~Paredes {\it et al.}, Nature~{\bf 429}, 277-281 (2004).

\bibitem{Fertig05}
C.~D.~Fertig {\it et al.}, Phys. Rev. Lett.~{\bf 94}, 120403 (2005).

\bibitem{Cornish00}
S.~L.~Cornish {\it et al.}, Phys. Rev. Lett.~{\bf 85}, 1795 (2000).

\bibitem{Mueller04}
E.~J.~Mueller, Phys. Rev. A~{\bf 70}, 041603(R) (2004).

\bibitem{Pupillo06}
G.~Pupillo, C.~J.~Williams, and N.~V.~Prokof'ev, Phys. Rev. A~{\bf 73}, 013408
(2006).

\bibitem{Schneider06}
B.~I.~Schneider, L.~A.~Collins, and S.~X.~Hu, Phys. Rev. E~{\bf 73}, 036708
(2006).

\bibitem{Olshanii98}
M.~Olshanii, Phys. Rev. Lett.~{\bf 81}, 938 (1998).
%D.~S.~Petrov, G.~V.~Shlyapnikov, and J.~T.~M.~Walraven, Phys. Rev.
%Lett.~{\bf 85}, 3745 (2000).

\end{thebibliography}
\end{document}